\documentclass[11pt]{article}
\usepackage{amsmath}
\usepackage{amsfonts}
\usepackage{amssymb}
\usepackage{graphicx}

\def\bea{\begin{eqnarray}}
\def\eea{\end{eqnarray}}

\begin{document}
\begin{center}
\LARGE { \bf  Tachyon  Warm-Intermediate Inflationary Universe
Model in High Dissipative Regime
  }
\end{center}
\begin{center}
{\bf M. R. Setare\footnote{rezakord@ipm.ir} \\  V. Kamali\footnote{vkamali1362@gmail.com}}\\
 { Department of Science, Payame Noor
University, Bijar, Iran}
 \\
 \end{center}
\vskip 3cm

\begin{abstract}
We consider tachyonic warm-inflationary models in the context of
intermediate inflation. We derive the characteristics of this
model in slow-roll approximation and develop our model in two
cases, 1- For a constant dissipative parameter $\Gamma$. 2-
$\Gamma$ as a function of tachyon field $\phi$. We also describe
scalar and tensor perturbations for this scenario. The parameters
appearing in our model are constrained by recent observational
data. We find that the level of non-Gaussianity for this model is
comparable with non-tachyonic model.
\\

\end{abstract}

\newpage

\section{Introduction}
Big Bang model have many long-standing problems (horizon,
flatness,...). These problems are solved in the framework of the
inflationary universe models \cite{1-i}. Scalar field as a source
of inflation provides the causal interpretation of the origin of
the distribution of large scale structure and observed anisotropy
of cosmological microwave background (CMB) \cite{2-i}. In
standard models for inflationary universe, the inflation period
is divided into two regimes, slow-roll and reheating epochs. In
slow-roll period kinetic energy remains small compared to the
potential terms. In this period, all interactions between scalar
fields (inflatons) and  other fields are neglected and the
universe inflates. Subsequently, in reheating period, the kinetic
energy  is comparable to  the potential energy and inflaton
starts an oscillation around the minimum of the potential losing
their energy to other fields present in the theory. So, the
reheating is the end period of inflation.\\ In warm inflationary
models radiation production occurs during inflationary period and
reheating is avoided \cite{4}. Thermal fluctuations may be
obtained during warm inflation. These fluctuations could play a
dominant role to produce initial fluctuations which are necessary
for Large-Scale Structure (LSS) formation. So the density
fluctuation arises from thermal rather than quantum fluctuation
\cite{3-i}. Warm inflationary period ends when the universe stops
inflating. After this period the universe enters in radiation
phase smoothly \cite{4}. Finally, remaining inflatons or dominant
radiation fields created the matter components of the universe.\\
Friedmann-Robertson-Walker (FRW) cosmological models in the
context of string/M-theory have related to brane-antibrane
configurations \cite{4-i}. Tachyon fields associated with
unstable D-branes may be responsible for inflation in early time
\cite{5-i}. If the tachyon field start to roll down the
potential, then universe dominated by a new form of matter will
smoothly evolve from inflationary universe to an era which is
dominated by a non-relativistic fluid \cite{1}. So, we could
explain the phase of acceleration expansion (inflation) in term
of tachyon field.\\ In this work we would like to consider the
tachyon warm inflationary universe in the particular scenario
"intermediate inflation" which is denoted by scale factor
$a(t)=a_0\exp(At^f), 0<f<1$ \cite{6-i}. The expansion of this
model is faster than power-law inflation ($a=t^p; p>1$), but
slower than standard de sitter inflation ($a=\exp(Ht)$). In
non-tachyonic inflation, potential has the form
$\phi^{-4\frac{1-f}{f}}$, but in the warm-tachyon inflation we
will receive to this form of potential
$\phi^{-4\frac{f-1}{2f-1}}$. The Harrison-Zeldovich spectrum of
density perturbation (i.e. $n_s=1$ \cite{6-i}) for these two
cases are given by $f=\frac{2}{3}$. The study of intermediate
inflationary model is motivated by string/M-theory \cite{v1}. In
term of this theory, if the high order curvature corrections to
Einstein-Hilbert action are proportional to the Gauss-Bonnet(GB)
term,  we obtain a free-ghost action. The GB term is the leading
order of the $\alpha$ (inverse string tension) expansion to the
low-energy string effective action \cite{v2}. This kind of theory
is applied for study of initial singularity problem \cite{v3},
black hole solutions \cite{v4} and the late time universe
acceleration \cite{v5}. The coupling of GB with dynamical
dilatonic scalar field in $4D$ dark energy model leads to an
intermediate form for scale factor, where $f=\frac{1}{2}$,
$A=\frac{2}{8\pi G n}$
with a constant parameter "n" \cite{v1}. So, intermediate inflation model may be derived from an effective theory at low dimensions of a fundamental string theory.     \\
The tachyon inflation is a k-inflation model \cite{n-1} for
scalar field $\phi$ with a positive potential $V(\phi)$. Tachyon
potentials have two special properties, firstly a maximum of
these potential is obtained where $\phi\rightarrow 0$ and second
property is the minimum of these potentials is obtained where
$\phi\rightarrow \infty$. In our  intermediate model we find an
exact solution for tachyon field potential in slow-roll
approximation which have the form $V(\phi)\sim\phi^{-\beta}$.
This form of potential has the above properties of tachyon field
potential. This special form of potential is everlasting, so we
consider our model in the context of the warm inflation to bring
 an end for inflation period.
 We know that in the warm inflationary
models, dissipative effect arises from a friction term. This
effect could describe the mechanisms that scalar fields decay
into a thermal bath via its interaction with other fields and
finally warm inflation ends when the universe heats up to become
radiation dominant. After inflation period the universe smoothly
connected with the radiation Big Bang phase. Intermediate
inflation in standard gravity with the presence of tachyon field
has been considered in Ref.\cite{8}. Also,  the warm tachyon
inflation in standard gravity has been studied
 in \cite{3}. As far as, we know, a model in which warm tachyon inflation
in the context of intermediate inflation has not been yet
considered. \\The paper is organized as follows. In section II,
we give a brief review about tachyon warm inflationary universe.
In section III we consider high dissipative warm-intermediate
inflationary phase in two cases 1- A constant dissipative
parameter $\Gamma$. 2- dissipative parameter $\Gamma$ as a
function of tachyon field $\phi$. In this section we also,
investigate the cosmological perturbations. In section IV we
consider the possibility of having non-gaussianity for
warm-tachyon model. Finally in section V, we close by concluding
remarks.
\section{Tachyon warm  inflationary universe}
Tachyonic inflation model in a spatially flat Friedmann Robertson
Walker (FRW) is described by an effective fluid which is
recognized by energy-momentum tensor
$T^{\mu}_{\nu}=diag(-\rho_{\phi},p_{\phi},p_{\phi},p_{\phi})$
\cite{1}, where energy density $\rho_{\phi}$, and pressure for
the tachyon field are defined by
\begin{eqnarray}\label{1}
\rho_{\phi}=\frac{V(\phi)}{\sqrt{1-\phi^2}}~~~~~~~~\\
\nonumber p_{\phi}=-V(\phi)\sqrt{1-\phi^2},
\end{eqnarray}
respectively, where $\phi$ denotes the tachyon scalar field and
$V(\phi)$ is the effective scalar potential associated with the
tachyon field. Characteristics of any tachyon field potential are
$\frac{dV}{d\phi}<0$ and $V(\phi\rightarrow 0)\rightarrow 0$
\cite{2}. Friedmann equation for spatially flat universe and
conservation equation, in the warm tachyon inflationary scenario
are given by \cite{3}
\begin{eqnarray}\label{2}
 H^2=\frac{8\pi}{3m_p^2}[\rho_{\phi}+\rho_{\gamma}],
\end{eqnarray}

\begin{eqnarray}\label{3}
\dot{\rho}_{\phi}+3H(\rho_{\phi}+p_{\phi})=-\Gamma
\dot{\phi}^2\Rightarrow~~~~\frac{\ddot{\phi}}{1-\dot{\phi}^2}+3H\dot{\phi}+\frac{V'}{V}=-\frac{\Gamma}{V}\sqrt{1-\dot{\phi}^2}\dot{\phi},
\end{eqnarray}
and
\begin{eqnarray}\label{4}
 \dot{\rho}_{\gamma}+4H\rho_{\gamma}=\Gamma \dot{\phi}^2,
\end{eqnarray}
where $H=\frac{\dot{a}}{a}$ is the Hubble factor, $a$ is a scale
factor, $m_p$ represents the Planck mass and  $\rho_{\gamma}$ is
the energy density of the radiation. Dissipative coefficient
$\Gamma$ has the dimension $m_p^5$. In the above equations dots
$"^{.}"$ mean derivative with respect to time and prime $"'"$ is
derivative with respect to $\phi$. During the inflationary epoch
the energy density (\ref{1}) is the order of the potential
$\rho_{\phi}\sim V$. Tachyon energy density $\rho_{\phi}$ in this
era dominates over the energy density of radiation
$\rho_{\phi}>\rho_{\gamma}$. In slow-roll regime ($\dot{\phi}\ll
1$ and $\ddot{\phi}\ll(3H+\Gamma/V)\dot{\phi}$ \cite{4}) and when
warm inflation radiation production is quasi-stable
($\dot{\rho}_{\gamma}\ll 4H\rho_{\gamma},
\dot{\rho}_{\gamma}\ll\Gamma\dot{\phi}^2$) the equations
(\ref{2}), (\ref{3}) and (\ref{4}) are reduced to
\begin{eqnarray}\label{5}
H^2=\frac{8\pi}{3m_p^2}V,
\end{eqnarray}

\begin{eqnarray}\label{6}
3H(1+r)\dot{\phi}=-\frac{V'}{V},
\end{eqnarray}
\begin{eqnarray}\label{7}
\rho_{\gamma}=\frac{\Gamma\dot{\phi}^2}{4H},
  \end{eqnarray}
where $r=\frac{\Gamma}{3HV}$. The main problem of inflation
theory is how to attach the universe to the end of the inflation
period. One of the solutions of this problem is the study of
inflation in the context of warm inflation \cite{m1}. In this
model radiation is produced during inflation period where its
energy density is kept nearly constant. This is phenomenologically
fulfilled by introducing the dissipation term,$\Gamma$, in the
equation of motion as we have seen in Eq.(\ref{3}). This term
grants a  continuous energy  transfer from scalar field energy
into a thermal bath. In this article we consider high dissipation
regimen ($r\gg 1$) where the dissipation coefficient $\Gamma$ is
much greater than the $3HV$. High dissipative and weak
dissipative regime for non-tachyonic warm inflation have been
considered in Ref.s \cite{m1} and \cite{m2} respectively. The
main attention was given to the high dissipative regime. This
regime is the more difficult of the two, and when high
dissipative regime is understood, the weak dissipative regime
could follow. Also, in Refs.\cite{m2} and \cite{m3}, the study of
warm inflation in the context of quantum field theory has been
fulfilled in high dissipative regime. Dissipative parameter
$\Gamma$ could be a constant or a positive function of $\phi$ by
the second law of thermodynamics. In some works $\Gamma$ and
potential of inflation have the same form\cite{3}, \cite{m4}. In
Ref.\cite{3}, perturbation parameters for warm tachyon inflation
have been obtained where $\Gamma=\Gamma_0=const$ and
$\Gamma=\Gamma(\phi)=V(\phi)$. So, in this work we will study the
intermediate tachyon warm inflation in high dissipative regime
for these two cases.
\\
From Eqs. (\ref{6}) and (\ref{7}) in high dissipation regimen we
get
\begin{eqnarray}\label{8}
\rho_{\gamma}=\frac{m_p^2}{32\pi r}[\frac{V'}{V}]^2.
\end{eqnarray}
We introduce the slow-roll parameter
\begin{eqnarray}\label{9}
\epsilon=-\frac{\dot{H}}{H^2}=\frac{m_p^2}{16\pi
r}[\frac{V'}{V}]^2\frac{1}{V}.
\end{eqnarray}
Using Eqs. (\ref{8}) and (\ref{9}) in slow-roll regime
($\rho_{\phi}\sim V$) we find a relation between the energy
densities $\rho_{\phi}$ and $\rho_{\gamma}$ as
\begin{eqnarray}\label{10}
\rho_{\gamma}=\frac{\epsilon}{2}\rho_{\phi}.
\end{eqnarray}
The second slow-roll parameter $\eta$ is given by
\begin{eqnarray}\label{11}
\eta=-\frac{\ddot{H}}{H\dot{H}}\simeq\frac{m_p^2}{8rV}[\frac{V''}{V}-\frac{1}{2}(\frac{V'}{V})^2].
\end{eqnarray}
The warm inflationary condition $\ddot{a}>0$ may be obtained by
parameter $\epsilon$, satisfying the relation
\begin{eqnarray}\label{12}
\epsilon <1.
\end{eqnarray}
From above equation and Eq.(\ref{10}) the tachyon warm inflation
epoch could take place when
\begin{equation}\label{13}
\rho_{\phi}>2\rho_{\gamma}.
\end{equation}
Inflation period ends when $\epsilon\simeq 1$ which implies
\begin{eqnarray}\label{14}
\rho_{\phi}\simeq 2\rho_{\gamma}.
\end{eqnarray}
The number of e-folds is given by
\begin{eqnarray}\label{15}
N(\phi)=-\frac{8\pi}{m_p^2}\int_{\phi_1}^{\phi_2}\frac{V^2}{V'}r
d\phi,
\end{eqnarray}
where $\phi_1$ and $\phi_2$ denote the begining and the end
inflatons.

\section{Intermediate inflation}
In this section we consider high dissipative tachyon intermediate
inflation. In intermediate inflation the scale factor follows the
law
\begin{eqnarray}\label{16}
a(t)=a_0\exp(At^f) ~~~~~~0<f<1,
\end{eqnarray}
where $A$ is a positive constant with units $m_p^f$. We would
like to consider our model in two important cases \cite{3}. 1-
$\Gamma$ is a constant parameter. 2- $\Gamma$ is a function of
$\phi$ ($\Gamma=f(\phi)=V(\phi)$)
\subsection{$\Gamma=\Gamma_0=const$}
Using Eqs. (\ref{5}), (\ref{6}) and (\ref{16}) with
$\Gamma=constant$, we get tachyon scalar field $\phi$, and
potential $V(\phi)$ as
\begin{eqnarray}\label{17}
\phi=\sqrt{\frac{3m_p^2(1-f)(fA)^2}{2\pi\Gamma_0(2f-1)^2}}t^{\frac{2f-1}{2}},
\end{eqnarray}
and
\begin{eqnarray}\label{18}
V(\phi)=B_1\phi^{-\beta_1},
\end{eqnarray}
where
\begin{eqnarray}\label{19}
\beta_1=\frac{4-4f}{2f-1}~~~~~~~~~B_1=\frac{3m_p^2(fA)^2}{8\pi}(\frac{\Gamma_0
2\pi(2f-1)^2}{3m_p^2(1-f)(fA)^2})^{-\frac{\beta_1}{2}}.
\end{eqnarray}
The Hubble parameter in term of $\phi$ becomes
\begin{eqnarray}\label{20}
H(\phi)=\sqrt{\frac{8\pi B_1}{3m_p^2}}\phi^{-\frac{\beta}{2}}.
\end{eqnarray}
From equation (\ref{18}), if $f>\frac{1}{2}$, $V(\phi)$ has the
characteristic of tachyon field potential ($\frac{dV}{d\phi}<0$
and $V(\phi\rightarrow 0)\rightarrow 0$), also these potentials
do not have a minimum \cite{5}. Slow-roll parameters $\epsilon$
and $\eta$ in term of tachyon field $\phi$ are obtained from Eqs.
(\ref{9}) and (\ref{11}).
\begin{eqnarray}\label{21}
\epsilon =\sqrt{\frac{3m_p^2 B_1}{32\pi
\Gamma_0^2}}\beta_1^2\phi^{-(\frac{\beta}{2}+2)},~~~~~~
\end{eqnarray}

\begin{eqnarray}\label{22}
\eta =\sqrt{\frac{3m_p^2 B_1}{32\pi
\Gamma_0^2}}\beta_1(\beta_1+2)\phi^{-(\frac{\beta}{2}+2)},
\end{eqnarray}
respectively. From Eq.(\ref{15}), the number of e-folds between
two fields $\phi_1=\phi(t_1)$ and $\phi_2=\phi(t_2)$
($\phi_1>\phi_2$ if $f>\frac{1}{2}$) is given by
\begin{eqnarray}\label{23}
N(\phi)=-\frac{8\pi}{m_p^2}\int_{\phi_1}^{\phi_2}\frac{V^2}{V'}\frac{\Gamma_0}{3HV}d\phi=\frac{2\Gamma_0}{\beta_1(\beta_1+4)}\sqrt{\frac{8\pi}{3m_p^2
B_1}}(\phi_2^{\beta_1/2+2}-\phi_1^{\beta_1/2+2})
\end{eqnarray}
$\phi_1$ is obtained at the begining of inflation ($\epsilon=1$),
$\phi_1^{-(\beta_1/2+2)}=\sqrt{\frac{32\pi \Gamma_0^2}{3m_p^2
B_1^2}}\frac{1}{\beta_1^2}$, so we can determine the value of
$\phi_2$ in term of $N$, $A$ and $f$. At this stage we consider
scalar and tensor perturbations for our model. For the tachyon
field in warm inflationary universe (in slow-roll and high
dissipative regime) the power spectrum of the curvature
perturbation and amplitude of tensor perturbation (which would
produce gravitational waves during inflation) are given by
\cite{3}
\begin{eqnarray}\label{24}
P_R\simeq\frac{\sqrt{3}}{30\pi^2}\exp(-2\Im(\phi))[(\frac{1}{\epsilon})^3\frac{9m_p^4}{128\pi^2r^2\sigma
V}]^{\frac{1}{4}}
\end{eqnarray}
\begin{eqnarray}\label{25}
P_T=\frac{16\pi}{m_p^2}(\frac{H}{2\pi})^2\coth[\frac{k}{2T}]\simeq\frac{32V}{3m_p^4}\coth[\frac{k}{2T}],
\end{eqnarray}
respectively. Temperature $T$ in extra factor
$\coth[\frac{k}{2T}]$ denotes, the temperature of the thermal
background of gravitational wave \cite{6} and
\begin{eqnarray}\label{26}
\Im(\phi)=-\int[\frac{1}{3Hr}(\frac{\Gamma}{V})'+\frac{9}{8}\frac{V'}{V}[1-\frac{(\ln
\Gamma)'(\ln V)'}{36H^2r}]]d\phi.
\end{eqnarray}
For $r\gg 1$ from Eqs.(\ref{24}) and (\ref{25}) tensor-scalar
ration is obtained as
\begin{eqnarray}\label{27}
R(k_0)\approx
\frac{240\sqrt{3}}{25m_p^2}[\frac{r^{\frac{1}{2}}\epsilon
H^3}{T_r}\exp[2\Im(\phi)]\coth(\frac{k}{2T})]|_{k=k_0}.
\end{eqnarray}
$R$ is important parameter. We can use the seven-year  Wilkinson
Microwave Anisotropy Probe (WMAP7) observations to find an upper
bound for $R$, from these results  we have $P_R\simeq 2.28\times
10^{-9}, R=0.21<0.36$  \cite{2-i}. Spectral indices $n_g$ and
$n_s$ were calculated in \cite{3}
\begin{eqnarray}\label{28}
n_g=-2\epsilon~~~~~~~~~~~~~~~~~~~~~~~~~~~~~~~~~~~~~~~~,\\
\nonumber
n_s=1-[\frac{3}{2}\eta+\epsilon(\frac{2V}{V'}[2\Im'(\phi)-\frac{r'}{4r}]-\frac{5}{2})].
\end{eqnarray}
In intermediate inflation, we obtained these parameters in term
of tachyon field
\begin{eqnarray}\label{29}
n_g=-\sqrt{\frac{3m_p^2\beta_1^2}{8\pi\Gamma_0^2}}\beta_1^2\phi^{-(\beta_1/2+2)}~~~~~~~~~~~~~~~~~~~~~,\\
\nonumber
n_s=1-\frac{3}{2}\eta-\frac{9}{4}\epsilon=1+\frac{3}{8}(\beta_1-4)\sqrt{\frac{\beta_1^4
B_1}{\Gamma_0^2 k}}\phi^{-(\beta_1/2+2)}.
\end{eqnarray}
Since $\frac{1}{2}<f<1$ we obviously see that the
Harrison-Zeldovich spectrum  (i.e, $n_s=1$) occurs for $\beta_1=4$
or equivalently $f=\frac{2}{3}$ which  agrees with tachyon
intermediate inflation \cite{8} and non-tachyonic inflation model
\cite{6-i},\cite{m5}. $n_s>1$ is equivalent to $\beta_1>4$  and
$n_s<1$ is equivalent to $\beta_1<4$. Runing in the spectral
indices $\frac{dn_s}{d\ln{k}}$ is one of the important parameters
which could be obtained by WMAP7 data \cite{2-i}.
\begin{eqnarray}\label{30}
\alpha_s=\frac{dn_s}{d\ln
k}=-\frac{dn_s}{dN}=-\frac{d\phi}{dN}\frac{dn_s}{d\phi}=-\frac{3m_p^2(\beta^2-16)}{124\pi}\frac{V'}{V^2r}\sqrt{\frac{\beta_1^4
B_1}{\Gamma_0^2 k}}\phi^{-(\beta_1/2+2)}.
\end{eqnarray}
 In terms of WMAP7 data, $\alpha_s$ is  approximately  $-0.038$ \cite{2-i}.
In the next subsection we will study the specific case in which
the dissipative  parameter is a function of $\phi$.
\subsection{$\Gamma=\Gamma{(\phi)}=V(\phi)$}
With $\Gamma=V(\phi)$ \cite{3},  and using the Eqs.(\ref{4}) and
(\ref{5}), the slow-roll parameters become
\begin{eqnarray}\label{31}
\epsilon=\frac{\sqrt{3}m_p\beta_2^2}{2\sqrt{8\pi B_2}}\phi^{\beta_2/2-2},~~~~~~\\
\nonumber \eta=\frac{\sqrt{3}m_p\beta_2(\beta_2+2)}{2\sqrt{8\pi
B_2}}\phi^{\beta_2/2-2},
\end{eqnarray}
where
\begin{eqnarray}\label{32}
\beta_2=-4(f-1),~~~~~~~~~B_2=\frac{3m_p^2(fA)^2}{8\pi
(2-2f)^{4f-4}},~~~~~~~~\phi=\sqrt{2(1-f)t}.
\end{eqnarray}
The Hubble parameter and potential in this case are given by
\begin{eqnarray}\label{33}
H(\phi)=\sqrt{8\pi
B_2/3m_p^2}\phi^{-\beta_2/2},~~~~~~~~~~V(\phi)=B_2\phi^{-\beta_2},
\end{eqnarray}
respectively. By using Eq.(\ref{15}) the number of e-folds is
obtained
\begin{eqnarray}\label{34}
N(\phi)=\sqrt{8\pi
B_2/3m_p^2}\int_{\phi_1}^{\phi_2}\phi^{-\beta_2/2+1}d\phi=\frac{2\sqrt{8\pi
B_2/3m_p^2}}{\beta_2+4}(\phi_2^{-\beta_2/2+2}-\phi_1^{-\beta_2/2+2}).
\end{eqnarray}
$\phi_1$ is obtained at the begining of inflation ($\epsilon=1$)
and the value of $\phi_2$ could be determined in term of $N$, $A$
and $f$. Important parameters $R(k_0)$, $n_s$ and $\alpha_s$
could be obtained in this case ($\Gamma=V$) as
\begin{eqnarray}\label{35}
R(k_0)=\frac{48\sqrt{3}}{5m_p^2}[(\frac{16(8\pi)^2\sigma^2\beta_2^12}{81(3m_p^2)^3B_2^8})]\phi^{\beta_2-\frac{7}{4}}\exp(\frac{\beta_2^3}{32\sqrt{8\pi
B_2/3mp^2}(\beta_2/2+2)}\phi^{-\beta_2/2-2})
\end{eqnarray}
\begin{eqnarray}\label{36}
n_s=1-\frac{3}{2}\eta+\frac{3}{4}\epsilon(1+\epsilon)
\end{eqnarray}
\begin{eqnarray}\label{37}
\alpha_s=-\frac{d\phi}{dN}\frac{dn_s}{d\phi}=\frac{3m_p^2\beta_2^2}{8\pi
B_2}(\frac{3\sqrt{3}m_p}{8\sqrt{2\pi
B_2}}\beta_2(\beta_2-4)\phi^{\frac{3}{2}\beta_2-5}-(\frac{3}{4}\beta_2+3)(\frac{\beta_2}{2}-2)\phi^{\beta_2-4})
\end{eqnarray}
respectively. These parameters may be constrained by WMAP7 data
\cite{2-i}.
\section{Non-Gaussianity}
In this section we consider the possibility of having
non-gaussianity for our model. So, we will obtain the bispectrum
of the gravitational field fluctuations which are generated
during warm-tachyon inflation period. The zero mode equation of
motion of tachyon field $\phi$ have been obtain in Eq.(\ref{3}).
This equation in slow-roll approximation and high-dissipative
regime converts to the equation
\begin{eqnarray}\label{38}
\frac{d\phi}{dt}=-\frac{1}{\Gamma}\frac{dV}{d\phi}
\end{eqnarray}
In order to study the fluctuation of the thacyon field
$\delta\phi(x,t)$, we need a local form of the above temporal or
non-local equation \cite{9}. We obtain this extended equation by
imposing a near-thermal-equilibrium Markovian approximation.
Therefore the fluctuation-dissipation theorem may be applied.
From this the tachyon field is described by stochastic system
which evolves according to a Langevin equation \cite{10}.
\begin{eqnarray}\label{39}
\frac{d\phi}{dt}=\frac{1}{\Gamma
u/V}[a^{-2}\nabla^2\phi-\frac{V'}{V}+\frac{u\eta}{V}]
\end{eqnarray}
where $\eta(x,t)$ is a stochastic source, and
$u=\sqrt{1+\nabla_{\mu}\phi\nabla^{\mu}\phi}$. Using the
fluctuation-dissipation theorem, the source term in momentum
space has these properties
\begin{eqnarray}\label{40}
<\eta>=0~~~~~~~~~~~~~~~~~~~~~~~~~~~~~~~~~~~~~~~~~~~~~~~~~~~~\\
\nonumber <\eta(\textbf{k},t)\eta(\textbf{k}',t')>=2\Gamma
T(2\pi)^3\delta^{(3)}(\textbf{k}-\textbf{k}')\delta(t-t')
\end{eqnarray}
Following the method which is used in Ref.\cite{9}, the full
tachyon field is expressed as
$\phi(x,t)=\phi_0(t)+\delta\phi(x,t)$, where $\phi_0$ is
homogeneous background field and $\delta\phi\ll\phi_0$. In order
to obtain the three-point correlation function of the density
perturbation in Fourier space or the bispectrum, we consider the
evolution equation  up to second order fluctuations
\begin{eqnarray}\label{41}
\delta\phi=\delta_1\phi+\delta_2\phi
\end{eqnarray}
where $\delta_1\phi$ is in first order of $\delta\phi$ and
$\delta_2\phi$ is in second order of $\delta\phi$. From
Eq.(\ref{39}), the evolution equations of Fourier modes of first
and second order fluctuations are
\begin{eqnarray}\label{42}
\frac{d}{dt}(\delta_1\phi(\textbf{k},t))=\frac{1}{\Gamma}[-k^2V(\phi_0)\delta_1\phi(\textbf{k},t)-V''\delta_1\phi(\textbf{k},t)+\eta(\textbf{k},t)]~~\\
\nonumber
\frac{d}{dt}(\delta_2\phi(\textbf{k},t))=\frac{1}{\Gamma}[-k^2V(\phi_0)\delta_2\phi(\textbf{k},t)-V''\delta_2\phi(\textbf{k},t)~~~~~~~~~~~~~~~\\
\nonumber
-(\delta_1\phi(\textbf{k},t))^2(V'k^2+\frac{1}{2}V''')]~~~~~~~~~~~~~~~~~~~~~~~
\end{eqnarray}
where these equations are obtained in a time period
$t_n-t_{n-1}=\frac{1}{H}$ and we could obtain a complete solution
over longer time by piecewise construction \cite{9}. The solution
of the above equations are
\begin{eqnarray}\label{43}
\delta_1\phi(\textbf{k},t)=A(k,t-t_{n-1})\int_{t_{n-1}}^{t}dt'\frac{\eta(\textbf{k},t')}{\Gamma}A^{-1}(k,t'-t_{n-1})\\
\nonumber
+A(k,t-t_{n-1})\delta\phi_1(\textbf{k}e^{-H(t_n-t_{n-1})},t_{n-1})~~~~~~~~~~~~~~~~~~~~~~~~~
\end{eqnarray}
and
\begin{eqnarray}\label{44}
\delta\phi(\textbf{k},t)=A(k,t-t_{n-1})\int_{t_{n-1}}^{t}dt'B(k,t')~~~~~~~~~~~~~~~~~~~~~~~~~~~\\
\nonumber
[\int\frac{dp^3}{(2\pi)^3}\delta_1\phi(\textbf{p},t')\delta_1\phi(\textbf{k}-\textbf{p},t')]A^{-1}(k,t'-t_{n-1})\\
\nonumber
+A(k,t-t_{n-1})\delta_2\phi(\textbf{k}e^{-H(t_n-t_{n-1})},t_{n-1})~~~~~~~~~~~~~
\end{eqnarray}
where
\begin{eqnarray}\label{45}
A(k,t)=\exp[-\int_{t_0}^{t}(\frac{k^2V(\phi_0(t'))}{\Gamma}+\frac{V''(\phi_0(t'))}{\Gamma})dt']~~~~~~~~~~~~~~~~~~~~~~\\
\nonumber
B(k,t)=-\frac{2V'(\phi_0(t))k^2+V'''(\phi_0(t))}{\Gamma}~~~~~~~~~~~~~~~~~~~~~~~~~~~~~~~~~~
\end{eqnarray}
The second term in solutions (\ref{43}) and (\ref{44}) are memory
terms \cite{12}. The important concept of freeze-out is given by
the relevance of these memory terms within the Hubble time. The
freeze-out momentum $k_F$ is defined  in momentum space as: When
$k<k_F$  the memory term remainds during the Hubble time and for
$k\geq k_F$ this term damps away. By definition $k_F$ have a
condition
\begin{eqnarray}\label{46}
\frac{V(\phi_0)k^2+V''(\phi_0)}{\Gamma H}>1
\end{eqnarray}
In warm-inflation we have $V''(\phi_0)<\Gamma H$ and above
condition may be simplified by this definition
\begin{eqnarray}\label{47}
k_F=\sqrt{\frac{\Gamma H}{V(\phi_0)}}
\end{eqnarray}
By using the  solutions (\ref{43}) and (\ref{44}), the
three-point function of the  fluctuations at the large scale
(when $t\simeq t_{60}$ or when the number of e-fold is equal to
60) will be computed before the end of inflation
\begin{eqnarray}\label{48}
<\delta\phi(\textbf{k}_1,t)\delta\phi(\textbf{k}_2,t)\delta\phi(\textbf{k}_3,t)>=A(k_3,t-t_{60}+1/H)~~~~~~~~~~~~~~~~~~~~~~~~~~~~~~~~~~~\\
\nonumber
\int_{t_{60}-1/H}^{t_{60}}A^{-1}(k_3,t'-t_{60}+1/H)B(k_3,t')~~~~~~~~~~~~~~~~~~~~~~~~~~~~~~~~~~~~~~~~~~~~~~~~~~\\
\nonumber
[\int\frac{dp^3}{(2\pi)^3}<\delta_1\phi(\textbf{k}_1,t)\delta_1\phi(\textbf{p},t')><\delta_1\phi(\textbf{k}_2,t)\delta_1\phi(\textbf{k}_3-\textbf{p},t')>]~~~~~~~~~~~~~~~~~~~~~~\\
\nonumber
A(k_3,t-t_{60}+1/H)<\delta_1\phi(\textbf{k}_1,t_{60})\delta_1\phi(\textbf{k}_2,t_{60})\delta_2\phi(\textbf{k}_3e^{-1},t_{60}-1/H)>~~~~~\\
\nonumber
+(\textbf{k}_1\leftrightarrow\textbf{k}_3)+(\textbf{k}_2\leftrightarrow\textbf{k}_3)~~~~~~~~~~~~~~~~~~~~~~~~~~~~~~~~~~~~~~~~~~~~~~~~~~~~~~~~~~~~~~~~~~
\end{eqnarray}
In the freeze-out region ($k>k_F$) and by using the
approximations $A\simeq 1$ and $B(k,t)\approx B(k_F,t_F)$ the
three-point function becomes
\begin{eqnarray}\label{49}
<\delta\phi(\textbf{k}_1,t)\delta\phi(\textbf{k}_2,t)\delta\phi(\textbf{k}_3,t)>\approx B(k_F,t_F)\Delta t_F~~~~~~~~~~~~~~~~~~~~\\
\nonumber
[\int\frac{dp^3}{(2\pi)^3}<\delta_1\phi(\textbf{k}_1,t)\delta_1\phi(\textbf{p},t')><\delta_1\phi(\textbf{k}_2,t)\delta_1\phi(\textbf{k}_3-\textbf{p},t')>\\
\nonumber
+(\textbf{k}_1\leftrightarrow\textbf{k}_3)+(\textbf{k}_2\leftrightarrow\textbf{k}_3)]~~~~~~~~\Delta
t_F=t_H-t_F\approx\frac{1}{H}\ln(\frac{k_F}{H})
\end{eqnarray}
where $t_F$ is the time when the  last three wavevectors
thermalizes, and $t_H$ is the time at Hubble crossing of the
smallest inflation perturbation modes. The bispectrum for
slow-roll, single field is given by
\begin{eqnarray}\label{50}
<\Phi(\textbf{k}_1\textbf{k}_2\textbf{k}_3)>=2f_{NL}(2\pi)^3\delta^3(\textbf{k}_1+\textbf{k}_2+\textbf{k}_3)[P_{\Phi}(\textbf{k}_1)P_{\Phi}(\textbf{k}_2)+perms]
\end{eqnarray}
where the gravitational field $\Phi$ has the simple form
\begin{eqnarray}\label{51}
\Phi(\textbf{k})=-\frac{3}{5}\frac{H}{\dot{\phi}}\delta\phi(\textbf{k})
\end{eqnarray}
So, $f_{NL}$ for warm-tachyon inflation in high dissipative
regime is
\begin{eqnarray}\label{52}
f_{NL}=-\frac{5}{3}(\frac{\dot{\phi}}{H})[\frac{1}{H}\ln(\frac{k_F}{H})(\frac{V'''(\phi_0(t_F))+2k_F^2V'(\phi_0(t_F))}{\Gamma})]
\end{eqnarray}
In Ref.\cite{9} the $f_{NL}$ parameter for non-tachyonic warm
inflation is obtained as
\begin{eqnarray}\label{53}
f_{NL}=-\frac{5}{3}(\frac{\dot{\phi}}{H})[\frac{1}{H}\ln(\frac{k_F}{H})\frac{V'''(\phi_0(t_F))}{\Gamma}]
\end{eqnarray}
For the same potential, we see the non-gaussianity in the
curvature  in warm-tachyon inflation (\ref{52}) is comparable to
the non-tachyonic warm-inflation (\ref{53}).
\section{Conclusion}
The study of inflationary epoch with intermediate scale factor
leads to overlasting form of potential ($\phi^{-\beta}$) which
agrees with tachyon potential properties and the study of warm
inflation model as a mechanism that gives an end for tachyon
inflation are motivated us to consider the tachyon
warm-intermediate inflation model.
 So in this paper we have investigated  the warm-tachyon-intermediate
inflationary models. We have studied this scenario in two
different cases of the dissipative coefficient $\Gamma$. Our
model have been described for $\Gamma=\Gamma_0=const$ and for
$\Gamma$ as a function of tachyon field $\phi$, i.e.
$\Gamma=f(\phi)=V(\phi)$. For these cases we have found exact
solutions of Friedmann  for spatially flat universe containing
tachyon scalar field $\phi(t)$. We also have obtained scalar
potential and Hubble parameter as a function of tachyon field. In
slow-roll approximation, explicit expressions for the
tensor-scalar ratio $R$, scalar spectrum index $n_s$ and its
running $\alpha_s$ have been obtained, we have constrained these
parameters by WMAP7 observational data. In $\Gamma=\Gamma_0$ case,
we introduced scalar field potential as $V(\phi)\propto
\phi^{-4\frac{f-1}{2f-1}}, 0<f<1$, but in non-tachyonic inflation
potential has the form $\phi^{-4\frac{1-f}{f}}$. Tachyonic
condition for potential $V(\phi)$ ($\frac{dV}{d\phi}<0$)
constrained $f$ as $\frac{1}{2}<f<1$. In this case, it is
possible in the slow-roll approximation to have the
Harrison-Zeldovich spectrum of density perturbation (i.e. $n_s=1$
\cite{6-i}), provided $f$ takes the value of $\frac{2}{3}$ which
 agrees with regular inflation model with a canonical scalar
field characterized by a quasi-exponential expansion.
 We found The non-Gaussianity of warm tachyon model is comparable with
 non-tachyonic warm inflation.

\end{document}